\RequirePackage{amsmath}
\documentclass[english]{eptcs}
\usepackage[T1]{fontenc}
\usepackage[utf8]{inputenc}
\usepackage{babel}
\usepackage{xargs}
\usepackage{amssymb}
\usepackage{amsthm}
\usepackage{stmaryrd}
\usepackage{algorithmic}
\usepackage{algorithm}
\usepackage{paralist}
\usepackage{booktabs}
\usepackage{wrapfig}
\usepackage{colortbl}
\usepackage{tabularx}
\usepackage{caption}
\usepackage{booktabs}
\usepackage{listings}
\newcolumntype{C}{>{\centering\arraybackslash}X}
\usepackage{tikz}
\usetikzlibrary{shapes,decorations.pathreplacing,calc,patterns,arrows,positioning,automata}

\newcommand{\singleton}{Singleton}
\newcommand{\mvc}{Model-View-Controller}
\newcommand{\broker}{Broker}
\newcommand{\mc}[1][0]{\ifthenelse{#1=1}{Model checking}{model checking}}
\newcommand{\nusmv}{NuSMV}

\newcommand{\fig}[2][0]{\ifthenelse{#1=1}{Figure~\ref{#2}}{Fig.~\ref{#2}}}
\newcommand{\tab}[2][0]{\ifthenelse{#1=1}{Table~\ref{#2}}{Tab.~\ref{#2}}}
\newcommand{\sect}[2][0]{\ifthenelse{#1=1}{Section~\ref{#2}}{Sect.~\ref{#2}}}
\newcommand{\mycite}[3][1]{\ifthenelse{#1=1}{#2 et al.~\cite{#3}}{#2~\ref{#3}}}

\newcommand{\dtforall}[1]{\ensuremath{\forall #1\colon}}

\newcommand{\caFActive}[1]{\Vert #1 \Vert}
\newcommand{\caFConn}[4]{#1.#2\rightarrow#3.#4}
\newcommand{\caTVal}[2]{#1.#2}

\newcommand{\ctglobally}[1]{\ensuremath{\Box #1}}
\newcommand{\cteventually}[1]{\ensuremath{\diamondsuit #1}}

\newtheorem{definition}{Definition}
\newtheorem{exmp}[definition]{Example}

\newcommand{\inlineeqno}[1]{\refstepcounter{equation}\label{#1}\ (\theequation)}

\newenvironment{istmp}[7][250]{
	\def\func{50}
	\def\fvar{1}
	\def\iwidth{#1}
	\newcommand{\isport}[2]{%
		\node[anchor=base west, text width=0.5*\iwidth pt] at (0,-\func-5 pt) {##1$\colon$};
		\node[anchor=base west, text width=0.5*\iwidth pt] at (0.5*\iwidth pt,-\func-5 pt) {##2};
		\pgfmathparse{\func+10}%
		\edef\func{\pgfmathresult}
	}
	\newcommand{\aline}[1][solid]{%
		\draw[##1] (0,-\func pt) -- (\iwidth pt,-\func pt);
		\pgfmathparse{\func+5}%
		\edef\func{\pgfmathresult}		
	}
	\newcommand{\var}[2]{%
		\node[anchor=base west, text width=5cm] at (0,-\func-5 pt) {\ifthenelse{\equal{\fvar}{1}}{\textbf{var}}{}};
		\node[anchor=base west, text width=5cm] at (1cm,-\func-5 pt) {##1$\colon$};
		\node[anchor=base west, text width=4cm] at (2cm,-\func-5 pt) {##2};		
		\pgfmathparse{\fvar+1}%
		\edef\fvar{\pgfmathresult}
		\pgfmathparse{\func+10}%
		\edef\func{\pgfmathresult}
	}
	\newcommand{\axiom}[2][]{%
		\node [anchor=base west, text width=5cm] at (0,-\func-8 pt) {##2};
		\ifthenelse{\equal{##1}{}}{}{
			\node [anchor=base west, text width=1cm] at (\iwidth pt-1cm,-\func-8 pt) {\inlineeqno{##1}};}
		\pgfmathparse{\func+15}
		\edef\func{\pgfmathresult}			
	}
	
	\begin{tikzpicture}	
	\draw[thick] (0,0) -- (\iwidth pt,0);
	\node [anchor=base west, text width=\iwidth] at (0,-10pt) {\textbf{ISpec} #2};
	\node [align=right, anchor=base east, text width=0.7*\iwidth] at (\iwidth pt,-10 pt) {\ifthenelse{\equal{#6}{}}{}{\textbf{based on} #6}
		\ifthenelse{\equal{#7}{}}{}{\textbf{uses} #7}};
	\draw [double,double distance=2pt] (0,-15pt) -- (\iwidth pt,-15pt);
	\node [anchor=base west, text width=\iwidth pt] at (0,-25pt) {\textbf{loc}: #3};
	\node [anchor=base west, text width=\iwidth pt] at (0,-35pt) {\textbf{in}: #4};
	\node [anchor=base west, text width=\iwidth pt] at (0,-45pt) {\textbf{out}: #5};
}
{
	\draw[thick] (0,-\func pt) -- (\iwidth pt,-\func pt);
	\end{tikzpicture}		
}

\newenvironment{ctstmp}[3][200]{
	\def\func{20}
	\def\cwidth{#1}
	\def\bfirst{0}
	\newcommand{\ctvar}[2]{%
		\node [anchor=base west, text width=1cm] at (0,-\func-5 pt) {\ifthenelse{\bfirst=0}{\textbf{var}}{}};
		\node[anchor=base west, text width=3cm] at (1cm,-\func-5 pt) {##1$\colon$};
		\node[anchor=base west, text width=3cm] at (3.5cm,-\func-5 pt) {##2};
		\edef\bfirst{1}	
		\pgfmathparse{\func+10}	
		\edef\func{\pgfmathresult}
	}
	
	\newcommand{\ctline}[1][solid]{%
		\draw[##1] (0,-\func pt) -- (\cwidth pt,-\func pt);
		\pgfmathparse{\func+5}
		\edef\func{\pgfmathresult}		
	}	
	\newcommand{\ctspace}[1][10]{%
		\pgfmathparse{\func+##1}		
		\edef\func{\pgfmathresult}	
	}		
	\newcommand{\ctaxiom}[2][]{%
		\node [anchor=base west, text width=\cwidth pt-1cm] at (0,-\func-8 pt) {##2};
		\ifthenelse{\equal{##1}{}}{}{
			\node [anchor=base west, text width=1cm] at (\cwidth pt-1cm,-\func-8 pt) {\inlineeqno{##1}};}
		\pgfmathparse{\func+15}		
		\edef\func{\pgfmathresult}			
	}
	\begin{tikzpicture}	
 	\draw [thick] (0,0) -- (\cwidth pt,0);
	\node [anchor=base west, text width=0.6*\cwidth] at (0,-10pt) {\textbf{Spec} #2};
	\node [align=right, anchor=base east, text width=0.4*\cwidth] at (\cwidth pt,-10pt) {\textbf{uses} \textit{#3}};	
	\draw [thin, double,double distance=2pt] (0,-15pt) -- (\cwidth pt,-15pt);
}
{
	\draw [thick] (0,-\func pt) -- (\cwidth pt,-\func pt);
	\end{tikzpicture}		
}

\title{Verifying Patterns of Dynamic Architectures \\ Using Model Checking%
\thanks{This work was partially funded by the German Federal Ministry of Education and Research (BMBF) under grant 01Is16043A.}}

\author{Diego Marmsoler \qquad Silvio Degenhardt\institute{Technical University of Munich\\Germany}\email{\{diego.marmsoler~|~silvio.degenhardt\}@tum.de}}
\def\titlerunning{Model Checking Patterns of Dynamic Architectures}
\def\authorrunning{D. Marmsoler, S. Degenhardt}

\makeatletter
\def\namedlabel#1#2{\begingroup
	#2%
	\def\@currentlabel{#2}%
	\phantomsection\label{#1}\endgroup
}
\makeatother
\begin{document}
\maketitle
\begin{abstract}
Architecture patterns capture architectural design experience and provide abstract solutions to recurring architectural design problems.
They consist of a description of component types and restrict component connection and activation.
Therefore, they guarantee some desired properties for architectures employing the pattern.
%
Unfortunately, most documented patterns do not provide a formal guarantee of whether their specification indeed leads to the desired guarantee.
%
Failure in doing so, however, might lead to wrong architectures, i.e., architectures wrongly supposed to show certain desired properties.
Since architectures, in general, have a high impact on the quality of the resulting system and architectural flaws are only difficult, if not to say impossible, to repair, this may lead to badly reparable quality issues in the resulting system.
%
To address this problem, we propose an approach based on \mc{} to verify pattern specifications w.r.t. their guarantees.
%
In the following we apply the approach to three well-known patterns for dynamic architectures:
the \singleton{}, the \mvc{}, and the \broker{} pattern.
%
Thereby, we discovered ambiguities and missing constraints for all three specifications.
%
Thus, we conclude that verifying patterns of dynamic architectures using \mc{} is feasible and useful to discover ambiguities and flaws in pattern specifications.\looseness-1
\end{abstract}

\section{Introduction}\label{sec:intro}
Architecture patterns capture architectural design experience and are regarded as the ``Grand Tool'' for designing a software system's architecture~\cite{Taylor2009}.
Patterns for dynamic architectures are patterns for architectures in which components may appear and disappear and connections may change over time~\cite{Wermelinger2001,Fiadeiro2013,Broy2014}.\looseness-1

Usually, a pattern provides \emph{abstract} solutions to recurring architectural design problems.
The solution is usually a specification of component types, connection, and activation constraints.
Moreover, it guarantees an overall property for architectures employing them~\cite{Shaw1996,Buschmann1996}.
This property then leads to certain, desired quality aspects of the resulting software system.
Consider, for example, the \singleton{} pattern. It consists of a component type singleton which is supposed to behave as follows:
if a new component of this type is required to be activated, this is only done if there is not yet any active component of this type available.
The guarantee for the overall architecture is then, that at every point in time at most one component of this type is activated.
This guarantee, in turn, leads to reduced resource utilization, since memory usage is minimized.

Unfortunately, for most patterns, there is no formal guarantee that their specification indeed leads to the desired guarantee.
For example, there is no guarantee that the specification of the \singleton{} pattern~\cite{Gamma1994} leads indeed to an architecture in which there is only one active component of that type during the whole execution. Indeed, as shown later on, there are many hidden assumptions about the environment which are left implicit and which are required in order to satisfy the guarantee.
Thus, there is actually no guarantee that architectures employing the specification indeed fulfill the desired quality attribute of reduced resource utilization.

However, since architects rely on patterns when designing an architecture, this may lead to wrong architecture decisions.
Wrong architecture decisions, however, may strongly influence a software systems quality~\cite{Bratthall2000,Garlan2000} and are only difficult, if not impossible, to repair~\cite{Garlan2000,li2015}.
In security-related applications, for example, a good architecture may eliminate or mitigate up to 92\% of the most dangerous weaknesses~\cite{MITRE2011}.
Similarly, two conclusions of the Ariane 5 explosion were that it could have been avoided at design time using correct architecture specifications and a ``software architect'' position was requested for future projects~\cite{Dalmau1997}.
To put it in the words of Garlan~\cite{Garlan2000}: ``a poor architecture can lead to a disaster for the whole project.''

Thus, we propose a $5$ step approach based on \mc{} to verify patterns w.r.t. their claimed guarantees:
\begin{compactenum}
	\item Review current literature about a pattern.
	\item Identify and formalize the interfaces of the involved component types.
	\item Model the behavior of component types in terms of abstract state machines.
	\item Specify the pattern's guarantee in terms of temporal logic formulae.
	\item Verify the guarantee by applying \mc{}.
\end{compactenum}

\noindent
To evaluate the approach, we applied it to verify three contemporary patterns for dynamic architectures:
The \singleton{} pattern, the \mvc{} pattern, and the \broker{} pattern.

In the following paper we report on our experience of applying \mc{} to verify architecture patterns. Thus, the major contributions of this paper are as follows:
\begin{compactenum}
	\item We describe an approach to formally verify patterns of dynamic architectures.
	\item We show feasibility and usability of applying \mc{} to verify patterns for dynamic architectures.
	\item We provide (verified) formalizations of the \mvc{} pattern.
	\item We describe characteristic (verified) properties for \singleton{}, MVC and \broker{} pattern.
\end{compactenum}

\noindent
The paper is structured as follows:
In \sect{sec:background} we first provide some background information about the techniques used to formalize the pattern specifications as well as of \mc{} in general.
Then, in \sect{sec:approach} we describe the details of the approach, demonstrated by means of a running example.
In \sect{sec:results} we summarize our results obtained for the \singleton{}, the \mvc{}, and the \broker{} pattern.
In \sect{sec:discussion} we then report on our experience of applying \mc{} to pattern verification and critically discuss our approach.
Finally, we provide an overview of related work in \sect{sec:rw} and conclude the paper with a summary of major findings, a discussion of possible implications, and points to future work in \sect{sec:conc}.
\section{Background}\label{sec:background}
In the following we briefly describe the background of our work.
Therefore, we first introduce the formalism used to specify patterns for dynamic architectures.
Then, we briefly discuss the basic idea behind \mc{} in general and the \nusmv{} symbolic model checker specifically.

\subsection{Specifying Constraints of Dynamic Architectures}\label{sec:dacl}
Over the last decades, a series of so-called architecture description languages appeared to support in the formal specification of dynamic software architectures. %
Examples include the Chemical Abstract Machine~\cite{Inverardi1995}, Rapide~\cite{Luckham1995}, Darwin~\cite{Magee1996}, Wright~\cite{Allen1997} and its dynamic extension~\cite{Allen1998}, $\Pi$-ADL~\cite{Oquendo2004}, xADL~\cite{Dashofy2001}, and ACME~\cite{Garlan2003}. %
Around the same time, some approaches emerged to formalize architectural styles and patterns~\cite{Abowd1995,Moriconi1995,Penix1997,LeMetayer1998,Bernardo2000,Mehta2003}. %
Only recently, however, formal models of dynamic architectures emerged~\cite{Wermelinger2002,Fiadeiro2013,Broy2014} and specification techniques for properties of such architectures were developed~\cite{Marmsoler_DACL,Marmsoler2016}.\looseness-1

In the following, we briefly summarize the major concepts and notations of dynamic architectures found in these works. Moreover, we introduce some techniques used in their specification.

\subsubsection{A Model of Dynamic Architectures}
In this work, a dynamic architecture is modeled by a set of so-called Configuration Traces (CTs)~\cite{Marmsoler2016}.
A CT, in turn, is a sequence of Architecture Configurations (CNFs) which consist of a set of active components, valuations of their ports with messages, and connections between their ports.

\subsubsection{Specification Techniques}
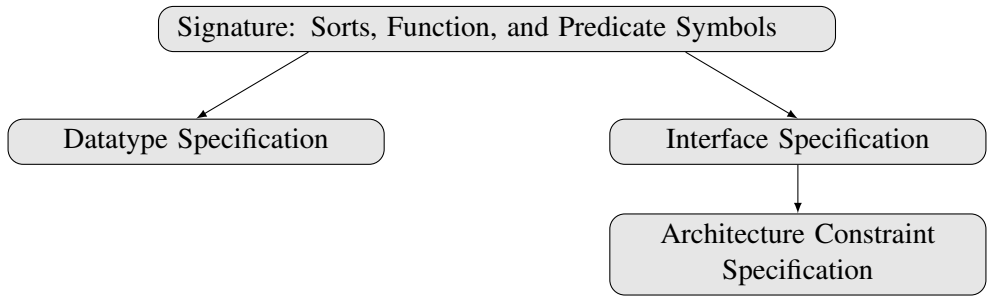
\begin{figure}[t]\centering
	\begin{tikzpicture}[phase1/.style={align=center, rectangle, draw, fill=black!10, rounded corners=0.2cm, minimum width=5cm, text width=4.5cm},phase2/.style={align=left, minimum width=4cm, text width=4cm},phase3/.style={align=left, minimum width=2.5cm, text width=2.5cm}]
	
	\node[phase1, text width=8.5cm, align=left, minimum width=9cm] (sig) at (2,4) {Signature: Sorts, Function, and Predicate Symbols};
	\node[phase1] (dt) at (-2,2.5) {Datatype Specification};
	\node[phase1] (if) at (6,2.5) {Interface Specification};
	\node[phase1] (ac) at (6,1) {Architecture Constraint \\ Specification};
	
	\draw[-latex] ($(sig.south west) + (2,0)$)--(dt.north);
	\draw[-latex] ($(sig.south east) - (2,0)$)--(if.north);	
	\draw[-latex] (if.south)--(ac.north);	
	\end{tikzpicture}
	\caption{Approach to specify properties of dynamic architectures.}
	\label{fig:PSL}
\end{figure}

For the specification of CTs, we employ algebraic specification techniques, interface specifications, and linear temporal logic~\cite{Manna2012}.
The overall approach is summarized in Fig.~\ref{fig:PSL}.

As a first step, a suitable signature is specified to introduce symbols for sets, functions, and predicates. These symbols form the primitive entities of the whole specification process. Datatype specifications and interface specifications as well as architecture constraint specifications are based on these symbols.

Then, datatypes are algebraically specified over the signature~\cite{Wirsing1990,Broy1996}. A Datatype Specification (DTS) consists of a set of so-called datatype assertions, built over datatype terms, to assert  characteristic properties of the datatype and provide meaning for the symbols introduced in the signature.

Interfaces are also directly specified over the signature. Therefore, a set of ports is typed by sorts of the corresponding signature by means of so-called port specifications. Then, an interface is specified by assigning an interface identifier with three sets of ports: local, input, and output ports. Finally, a set of interface assertions is associated with each interface identifier to specify component types, i.e., interfaces with associated global invariants.

Finally, architecture constraints can be specified by means of configuration trace assertions over the interfaces.
Configuration Trace Assertions (CTAs) are a temporal specification technique based on linear temporal logic~\cite{Manna2012} to specify sets of CTs.
They allow the specification of temporal properties over component interfaces.
Thereby, port names denote the valuation of a component port in a configuration and can be used as variables in algebraic terms as well. For example, $c.p$ denotes the current valuation of port $p$ of component $c$.
Moreover, CTAs allow for the specification of activation and connection predicates:
\begin{compactitem}
	\item \emph{Activation predicates} can be used to specify activation and deactivation of components. An activation of component $c$, for example, is denoted with $\caFActive{c}$.
	\item \emph{Connection predicates} can be used to specify connection between component ports. A connection between port $p$ of component $c$ and port $p'$ of component $c'$, for example, is denoted with $\caFConn{c}{p}{c'}{p'}$.
\end{compactitem}

\subsubsection{Configuration Diagrams}
We use Configuration Diagrams (CDs) as introduced in \cite{Marmsoler_DACL} as a graphical notation to support the specification of interfaces.
A CD is a graph whose nodes resemble interfaces (group of ports) and whose edges denote connections between component ports.
CDs can be annotated by certain common activation and connection constraints:
\begin{compactitem}
	\item \emph{Activation annotations} can be used to introduce common activation constraints, such as min./max. number of components of a certain type.
	\item \emph{Connection annotations} can be used to denote connection constraints, such as required connections between components of a certain type.
\end{compactitem}

\subsection{Model Checking}
\mc[1]{}~\cite{Baier:2008:PMC:1373322} (MC) is a technique for automatically verifying the correctness of certain properties against a model of a finite-state system.
The model of the system is thereby usually given in terms of a finite state machine and the properties are specified in terms of temporal logic formulae.

In this work we will use the \nusmv{} symbolic model checker~\cite{cavada2013nusmv}. A \nusmv{} model always consists of several modules, each of which represents a distinct state machine. Each module consists of 2  main parts:
a VAR part for declaring module variables and an ASSIGN part for defining the logic of the module, i.e., the initial value of the variables as well as state changes. Moreover, an optional part DEFINE can be used to define variable abbreviations.
Modules can interact with each other through parameters (input) and (output) variables. Every module has to specify the parameter it gets, which acts like input variables and can itself define variables in the DEFINE section that can be read by other modules in their transfer parameters.
Properties can then be expressed in LTL as well as CTL and checked against the model.
\section{Approach}\label{sec:approach}
\fig[1]{fig:structure} provides an overview of our approach to verify patterns for dynamic architectures and the corresponding artifacts:
\begin{figure}
	\centering 
	\tikzstyle{block} = [rectangle, draw, fill=black!10, text width=3cm, text centered, rounded corners, minimum height=1cm]
	\tikzstyle{line} = [draw, -latex']
	\tikzstyle{cloud} = [text width=3cm, text centered]	
	\begin{tikzpicture}[node distance = 2cm, auto]
	\node [block] (LS) {Literature Study};
	\node [cloud, below right = 0.4cm and -22pt of LS] (IPS) {\footnotesize Informal \\ Pattern Specification};
	\node [block, right = 1cm of LS] (IS) {Interface Specification};
	\node [cloud, above right = 0.5cm and -2cm of IS] (FIS) {\footnotesize Formal Interface Specification};			
	\node [block, above right = 0pt and 1cm of IS] (BHV) {Model of the Pattern};
	\node [block, below right = 0pt and 1cm of IS] (CNS) {Specify Guarantees};
	\node [cloud, above right = 0.7cm and 5cm of IS] (FPS) {\footnotesize Formal Model Pattern};
	\node [cloud, below right = 0.7cm and 5cm of IS] (FCS) {\footnotesize Formal Specification Guarantee};	
	\node [block, right = 5cm of IS] (VER) {Verification};
	\path [line] (LS) -- (IS);
	\path [line] (IS) --++ (2cm,0) |- (BHV);
	\path [line] (IS) --++ (2cm,0) |- (CNS);
	\path [line] (BHV)--++ (2cm,0) |- (VER);
	\path [line] (CNS)--++ (2cm,0) |- (VER);
	\path [draw, dashed] (LS) --++ (1.7cm,0) -- (IPS);
	\path [draw, dashed] (IS) --++ (1.9cm,0) -- (FIS);	
	\path [draw, dashed] (BHV) --++ (1.7cm,0) --++ (0.8cm,0.3cm);
	\path [draw, dashed] (CNS) --++ (1.7cm,0) --++ (1.2cm,-0.5cm);	
	\end{tikzpicture}
	\caption{General approach}\label{fig:structure} 
\end{figure}
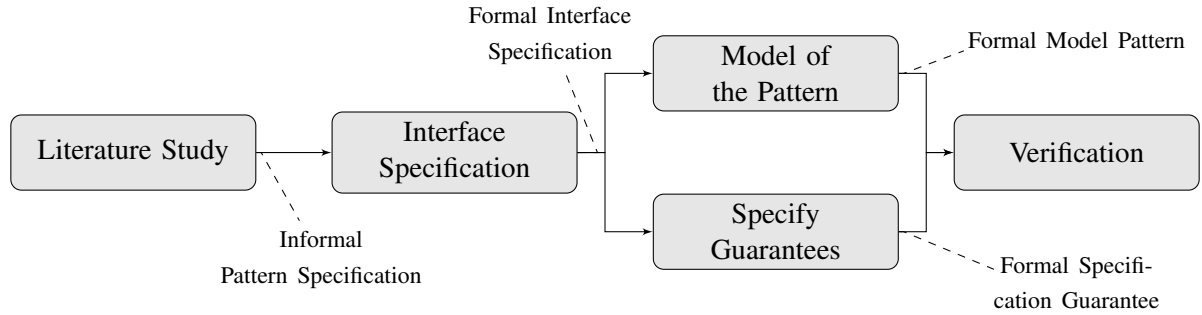
After consulting related literature describing a pattern, we identify and specify the interfaces of involved component types. Then, the logic of each component type is modeled by a state machine and the guarantee of the pattern is given in terms of configuration trace assertions.
Then, the model of the pattern as well as the formalized guarantee is translated into a corresponding \nusmv{} specification to verify whether the guarantees indeed hold.

\begin{exmp} [Running Example: \mvc{}]
	In the following, we demonstrate each step of the approach by means of a running example. Therefore, we choose the \mvc{} (MVC) patter which is commonly used for the design of human-computer interfaces~\cite{Buschmann1996,Shaw1996,Taylor2009}.
	The pattern is well-suited for demonstration since it is not to complex, yet it provides all important aspects of pattern specifications.
\end{exmp}

\subsection{Literature Study}\label{sec:ls}
The study of a pattern starts by consulting current literature to identify the major component types and constraints about their interrelationships and activation. Moreover, described claims about the guarantees of a pattern should be collected and documented.

\begin{exmp}[MVC Literature Study]\label{exmp:mvc:ls}
	It describes three types of components:
	\begin{description}
		\item[Model:] a unique component responsible for consistently storing the necessary data.
		\item[View:] displays the data of the model to the user.
		\item[Controller:] handles user inputs.
	\end{description}
	The pattern requires that there exists a designated controller component for each view component. A controller component recognizes user input (usually through events) and invokes a suitable service in the (unique) model component to update the model data accordingly. After this process, all views (and corresponding controllers) have to be updated with the new data from the model which is usually done through notifications.
	
	The claimed benefits of the MVC-pattern are described as follows:
	\begin{compactitem}
		\item[\namedlabel{benefit:b1}{B1}:] Single point of data storage.
		\item[\namedlabel{benefit:b2}{B2}:] Data consistency throughout the model and the views.
		\item[\namedlabel{benefit:b3}{B3}:] Easy to extend with new views.
		\item[\namedlabel{benefit:b4}{B4}:] High cohesion and low coupling of the components.
	\end{compactitem}
\end{exmp}

\subsection{Interface Specification}\label{sec:ispec}
Having an informal description of the pattern, we apply the techniques described in \sect{sec:dacl} to formalize the specification.
As a first step, the syntactic interfaces of the involved component types have to be formalized. As described in \sect{sec:dacl}, this can be done either graphically by means of configuration diagrams, or it can be done using interface specification templates. For each interface we specify local, input and output ports as well as their types.

\begin{exmp}[MVC Interface Specification]\label{exmp:ispec}
	As already described in the informal description of the pattern, the pattern consists of three types of components: Model, View, and Controller components.

	\noindent
	\begin{minipage}{\textwidth}
		\begin{wrapfigure}{r}{0.5\textwidth}
		\textnormal{			
			\centering
			\begin{istmp}[220]{Model}{service\_ele,~data}{service,~getData}{output,~notificate}{}{Array}
				\aline[dashed]		
				\isport{service, service\_ele}{$-1\dots2$}
				\isport{data, output}{array $0\dots2$ of $0\dots 10$}
				\isport{getData, notificate}{bool}										
			\end{istmp}
			\vspace{-15pt}
			\captionof{figure}{Model Interface Specification.\label{fig:ispec:mod}}
			\vspace{10pt}
			\begin{istmp}[220]{View}{view\_data,~view\_ele}{notification,~model\_data, element}{getData}{}{Array}
				\aline[dashed]		
				\isport{view\_data}{$0\dots 10$}
				\isport{view\_ele, element}{$0\dots 2$}
				\isport{model\_data}{array $0\dots2$ of $0\dots 10$}					
				\isport{notification, getData}{bool}	
			\end{istmp}
			\vspace{-15pt}
			\captionof{figure}{View Interface Specification.\label{fig:ispec:view}}		
			\vspace{10pt}		
			\begin{istmp}[220]{Controller}{event\_actual}{contr\_id,~random}{service}{}{}
				\aline[dashed]		
				\isport{event\_actual, service}{$-1\dots 2$}
				\isport{contr\_id, random}{$0\dots3$}
			\end{istmp}
			\vspace{-15pt}		
			\captionof{figure}{Controller Interface Specification.\label{fig:ispec:cnt}}}
		\end{wrapfigure}

		\fig[1]{fig:ispec:mod} provides a formal specification of the interface for Model components. Each Model has two local ports: \emph{data} to store all the data and \emph{service\_ele} to store the name of the currently running service.
		It receives a service call through the \emph{service} input port and a request to deliver the current data state through the \emph{getData} input port. Moreover, it provides its data through the \emph{output} output port and notifies its environment about data changes through its \emph{notificate} output port.

		In \fig{fig:ispec:view} we provide a formal specification of the interface for view components. A view component stores its data in a corresponding local port named \emph{view\_data}. Since we have more components of type view, each view has an associated identifier which is stored in local port \emph{view\_ele} and which is set at the beginning through its \emph{element} input port. Moreover, it gets notified about data updates and receives them through the \emph{notification} and \emph{model\_data} input ports. Finally, it requests new data from the model through its \emph{getData} output port.

	\fig[1]{fig:ispec:cnt} provides a specification of interfaces for controller components. The event a controller is currently working on is stored in the \emph{event\_actual} local port. Again, each controller has a unique identifier which it receives through the corresponding \emph{contr\_id} input port. Through its \emph{random} input port, a controller receives a notification about the occurrence of an event. The corresponding service invocation is then communicated through its \emph{service} output port.
	\end{minipage}	
\end{exmp}

\noindent
Having a formal specification of the components interfaces allows now to specify the behavior of a component type as well as the claimed guarantees over these interfaces.

\subsection{Model of the Pattern}\label{sec:model}
The behavior of the different component types is modeled using traditional mealy state machines. Thereby, each component type consists of a set of control states of which one is active at each point in time. State changes are triggered by valuations of a component's input ports and may result in valuations of component output ports as well as in a change of the currently active control state.

\begin{exmp}[MVC Behavior Specification]\label{exmp:model}
	In the following we specify the abstract behavior of model, view, and controller components, respectively.

	\noindent
	\begin{minipage}{\textwidth}
		\begin{wrapfigure}{r}{0.45\textwidth}
			\centering
			\begin{tikzpicture}[->,>=stealth',shorten >=1pt,auto,node distance=2.8cm, semithick,font=\footnotesize\normalfont]
				\tikzstyle{every state}=[fill=black!10,draw,text=black,text width=1.5cm,align=center,inner sep=0,outer sep=0,minimum width=1.8cm,font=\small\normalfont]
				
				\node[initial,state] (A) {update};
				\node[state] (B) [right= 1.5cm of A] {notifi-\\cation};
				
				\path (A) edge [bend left] node {(1,2)} (B)
				(B) edge [bend left] node {(3,4)} (A)
				(A) edge [loop above] node[align=center] {(5,6)} (A);
			\end{tikzpicture}
			\captionof{figure}{Model behavior.\label{fig:bhv:model}}
			
			\begin{tikzpicture}[->,>=stealth',shorten >=1pt,auto,node distance=2.8cm, semithick,font=\footnotesize\normalfont]
				\tikzstyle{every state}=[fill=black!10,draw,text=black,text width=1.5cm,align=center,inner sep=0,outer sep=0,minimum width=1.8cm,font=\small\normalfont]
	
				\node[initial,state] (A) {busy};
				\node[state] (B) [right= 1.5cm of A] {idle};

				\path (A) edge [bend left] node {(1)} (B)
				(B) edge [loop above] node[align=center] {(2)} (B)
				(B) edge [bend left] node {(3)} (A);
			\end{tikzpicture}
			\captionof{figure}{View behavior.\label{fig:bhv:view}}
			
			\begin{tikzpicture}[->,>=stealth',shorten >=1pt,auto,node distance=2.8cm, semithick,font=\footnotesize\normalfont]
				\tikzstyle{every state}=[fill=black!10,draw,text=black,text width=1.5cm,align=center,inner sep=0,outer sep=0,minimum width=1.8cm,font=\small\normalfont]

				\node[initial,state] (A) {control};

				\path (A) edge [loop above] node[align=center] {(1)} (A);
			\end{tikzpicture}
			\captionof{figure}{Controller behavior.\label{fig:bhv:controller}}				
		\end{wrapfigure}
		\fig[1]{fig:bhv:model} provides a specification of the behavior of model components.
		It is initialized (start) by setting \emph{service\_ele} to $-1$ and initializing its data store \emph{data}.
		If it gets a \emph{service}!=$-1$ it outputs a \emph{notificate}=false event, sets \emph{service\_ele}=\emph{service}, and changes to control state \emph{notification}. Thereby, depending on the valuation of its \emph{getData} input port, it returns either an empty output (1) or a copy of \emph{data} on its \emph{output} port (2).
		In control state \emph{notification}, a model immediately changes back to \emph{update}.
		Thereby it changes its data store \emph{data[service\_ele]} depending on the effects of the service call and notifies its environment about this changes by setting \emph{notificate}=true. Again, depending on the valuation of its \emph{getData} port it returns either an empty output (3) or a copy of \emph{data} on its \emph{output} port (4).
		Finally, in control state \emph{update} again, it waits for incoming data requests and depending on the valuation of its \emph{getData} port it returns either an empty output (5) or a copy of \emph{data} on its \emph{output} port (6).

		The behavior of components of type view is described by the state machine depicted in \fig{fig:bhv:view}.
		A view component is initialized (start) with a unique identifier, received through its \emph{element} input port and stored in the \emph{view\_ele} local port. It starts in control state \emph{busy} where it immediately changes to the \emph{idle} state, outputs a request to receive data by setting \emph{getData}=true, and updates its local data store accordingly by setting \emph{view\_data}=\emph{model\_data[view\_ele]} (1).
		It remains idle until it gets notified about a data-change from the model (2).
		After getting notified through its input port \emph{notification}, a view changes again its control state back to its initial state \emph{busy} (3).

		Finally, \fig{fig:bhv:controller} models the behavior of a controller component. Similar as a view, a controller gets initialized (start) by an identifier. It then continuously generates events which it forwards to the model by sending a corresponding service identifier through its output port \emph{service} (1).
	\end{minipage}
\end{exmp}

\subsection{Specify Constraints} \label{sec:constraints}
As a next step, one can start to formalize the claims identified in \sect{sec:ls} over the formal interface specification developed in \sect{sec:ispec}. Therefore, one can again leverage the methods and techniques presented in \sect{sec:background}.

\begin{exmp}[MVC Guarantee Specification]\label{exmp:mvc:const}
	In Ex.~\ref{exmp:mvc:ls} we identified several claims made about a system build in a MVC pattern. In the following we use a so-called configuration trace assertion template to formalize some of these guarantees of the MVC pattern.
	
	\noindent
	\begin{minipage}{\textwidth}
		\begin{wrapfigure}[10]{r}{0.60\textwidth}
			\centering			\vspace{-10pt}
			\textnormal{			
			\begin{ctstmp}[270]{MVC}{Array}
				\ctvar{$m$}{Model}
				\ctvar{$v$}{View}
				\ctvar{$c$}{Controller}		
				\ctline[dashed]
				\ctspace[5pt]		
				\ctaxiom[eq:mvc:1]{$\ctglobally{\Big(\caTVal{c}{\mathit{service}\neq-1} \implies \dtforall{v}\big(\cteventually{(\caTVal{v}{\mathit{getData}})}\big)\Big)}$}
				\ctspace[5pt]
				\ctaxiom[eq:mvc:2]{$\ctglobally{\Big(x=\caTVal{m}{\mathit{data}[\caTVal{v}{\mathit{view\_ele}}]} \implies \cteventually{\big(\caTVal{v}{\mathit{view\_Data}}=x\big)}\Big)}$}
				\ctspace[5pt]	
			\end{ctstmp}}
			\vspace{-15pt}
			\captionof{figure}{Specification of MVC guarantees.}
			\label{fig:spec_mvc} 
		\end{wrapfigure}
		\fig[1]{fig:spec_mvc} provides a formal specification of two properties for MVC Architectures.
		Eq.~\eqref{eq:mvc:1} requires that whenever a controller executes a service (in response to a user event), the views will at some point in the future request an update of their data.
		Eq.~\eqref{eq:mvc:2}, on the other hand, requires data consistency of the model and the views, i.e., that the data in the views is always updated according to their corresponding data in the model.\looseness-1
	\end{minipage}
\end{exmp}

\subsection{Verification} \label{sec:verification}
In the last step we apply \mc{} to verify the guarantees against the model of the pattern. First, the interface specification developed in \sect{sec:ispec} is used to build a corresponding \nusmv{} structure. Then, the model of the pattern developed in \sect{sec:model} is systematically translated to a corresponding \nusmv{} model. Finally, the specification of the pattern guarantees developed in \sect{sec:constraints} are transferred to a corresponding LTL specification in \nusmv{} and verified against the model.

\subsubsection{Translate Interfaces}
The specification of a pattern's interfaces is used to develop a raw structure of the \nusmv{} model. The general structure of a \nusmv{} module consists of a VAR, ASSIGN and DEFINE part. Algorithm~\ref{alg:temp} describes the systematic translation of a pattern's interface specification to a \nusmv{} raw structure.

\subsubsection{Translate Model}
In the next step, the \nusmv{} template is enhanced using the specification of component type behavior. Algorithm~\ref{alg:enhanced_temp} describes the systematic translation of a behavioral specification to a corresponding \nusmv{} pattern template.\looseness-1

\noindent
\begin{minipage}[t][7.5cm][t]{.48\linewidth}
\begin{algorithm}[H]
	\caption{Translate Interfaces to \nusmv{}}
	\label{alg:temp}
	\begin{algorithmic}[1]\small
		\REQUIRE interface specification of component types
		\STATE create a new \nusmv{}-File with one main module
		\FORALL {interface specifications \emph{is}}
			\STATE{create a new module for \emph{is}}
			\FORALL {input ports \emph{ip} of \emph{is}}
				\STATE{create module parameter for \emph{ip}}
			\ENDFOR
			\FORALL {local ports \emph{lp} of \emph{is}}
				\STATE{create module variable for \emph{lp}}
			\ENDFOR
			\FORALL {output ports \emph{op} of \emph{is}}
				\STATE{create variable in DEFINE part for \emph{op}}
			\ENDFOR
		\ENDFOR
		\RETURN	\nusmv{} template for the pattern
	\end{algorithmic}
\end{algorithm}
\end{minipage}
\hspace{10pt}
\begin{minipage}[t][7.5cm][t]{.48\linewidth}
	\begin{algorithm}[H]
		\caption{Translate Model to \nusmv{}}
		\label{alg:enhanced_temp}
		\begin{algorithmic}[1]\small
			\REQUIRE \nusmv{} template + model (state machines)
			\FORALL {component types \emph{ct}}
			\STATE create variable \emph{controlState} with entries for each control state in \emph{ct}'s VAR part
			\STATE define initial states of all local variables in \emph{ct}'s ASSIGN part
			\STATE encode state machine using \emph{next} statements in \emph{ct}s ASSIGN part
			\STATE define the valuation of output variables in the \emph{ct}'s DEFINE part
			\ENDFOR		
			\RETURN	enhanced \nusmv{} model with logic for all component type modules
		\end{algorithmic}
	\end{algorithm}
\end{minipage}



%

\noindent
Finally, the concrete architecture is configured in module main. First, components are instantiated and then, the configuration is encoded by connecting output ports (variables in DEFINE part) to corresponding input ports (module parameters) of the corresponding components.

\subsubsection{Translate Pattern Guarantees}
Now, as the \nusmv{} model is complete, we translate the guarantees formalized in \sect{sec:constraints} to corresponding \nusmv{} LTL specifications.

\begin{exmp}[Verifying the MVC pattern]
	\lstset{tabsize=2}
	\lstdefinelanguage{NuSMV}
	{morekeywords={MODULE,VAR,ASSIGN,DEFINE,LTLSPEC}}
	\lstset{language=NuSMV,basicstyle=\small\normalfont}
	Listing~\ref{lst:model} shows the \nusmv{} code resulting by applying Alg.~\ref{alg:temp} and Alg.~\ref{alg:enhanced_temp} to the specifications developed in Ex.~\ref{exmp:ispec} and Ex.~\ref{exmp:model}.

\begin{lstlisting}[frame=single,caption=\nusmv{} code for module \emph{model}.,label=lst:model,basewidth = {5.5pt}]
MODULE model (service, getData)
VAR
	controlState: {Update,Notification};
	service_ele : -1..2;
	data : array 0..2 of 0..10;
ASSIGN
	init(controlState) := Update;
	init (service_ele) := -1;
	init (data[0]) := 7;
	init (data[1]) := 5;
	init (data[2]) := 3;
	next (controlState) := case
		controlState=Update & service != -1 : Notification;
		TRUE : Update;
	esac;
	next(service_ele):= service;
		next(data[0]):=	case
		controlState = Notification & service_ele = 0 : (data[0] + 7) mod 10;
		TRUE : data[0];
	esac;
	next(data[1]):=	case
		controlState = Notification & service_ele = 1 : (data[1] + 5) mod 10;
		TRUE : data[1];
	esac;
	next(data[2]):=	case
		controlState = Notification & service_ele = 2 : (data[2] + 3) mod 10;
		TRUE : data[2];
	esac;
DEFINE
	output := 	case
		getData : data;
		TRUE : output_empty;
	esac;
	output_empty:= [-1,-1,-1]; 
	notifacte:= case
		controlState = Notification : TRUE;
		TRUE : FALSE;
	esac;
\end{lstlisting}
Similarly, the models for view and controller components are translated to \nusmv{}.

As a next step, the main module is build to coordinate the different modules. Listing~\ref{lst:MainModule} depicts the corresponding \nusmv{} code.
\begin{lstlisting}[frame=single,caption=\nusmv{} code for module \emph{main}.,label=lst:MainModule,basewidth = {5.5pt}]
MODULE main
VAR
	view1: view(model.notifacte,model.output,0);
	controller1: controller(1, random);
	view2: view(model.notifacte,model.output,1);
	controller2: controller(2, random);
	view3: view(model.notifacte,model.output,2);
	controller3: controller(3, random);
	model: model(service, getData);
	random: 0..3;
ASSIGN
	init(random) := 0;
	next(random) := {1,2,3};
DEFINE
	service := controller1.service + controller2.service + controller3.service + 2;
	getData := view1.getData | view2.getData | view3.getData;
\end{lstlisting}
	The main module is instantiating the components in the VAR part. It passes to every component the needed parameter. Most of them are output variables from the other components, but for example the IDs are passed as a static numeric. In the VAR part we also find a variable \emph{random}. The random variable is necessary to decide which of the controllers is allowed to invoke a service in the model. Without this restriction multiple controllers could invoke a service even though the model can only handle one call.\\
	In the DEFINE section we see a \emph{service} and a \emph{getData} variable. \emph{service} indicates which service is called and \emph{getData} whether a view is requesting an update of the data.
	
	Finally, the specification of the guarantees developed in Ex.\ref{exmp:mvc:const} is translated to a corresponding LTL specification in \nusmv{} (Listing~\ref{lst:mvc_ltl}).
\begin{lstlisting}[frame=single,caption=LTL Specification,label=lst:mvc_ltl,basewidth = {5.5pt}]
LTLSPEC G ((controller1.service + controller2.service + controller3.service)> -3)
	-> (F view1.getData) & (F view2.getData) & (F view3.getData)

LTLSPEC G model.data[view1.view_ele] = 0 -> F view1.view_data = 0
LTLSPEC G model.data[view2.view_ele] = 0 -> F view2.view_data = 0
...
\end{lstlisting}	
\end{exmp}
\section{Results}\label{sec:results}
We applied the approach to formalize and verify three contemporary patterns for dynamic architectures: The \singleton{}, the MVC, and the \broker{} pattern. \tab[1]{tab:res} provides an overview of our results\footnote{All the corresponding \nusmv scripts can be downloaded at \url{http://www.marmsoler.com/mc/}.}. In the following we discuss them in more detail.

\begin{table}\small
	\caption{Analysis Results.\label{tab:res}}
	\begin{tabular}{rlll}
		\toprule	
		\textbf{Pattern} & \textbf{Prop.} & \textbf{Type} & \textbf{Description} \\ 
		\midrule
		\emph{\singleton{}} & \namedlabel{res:s:1}{S1} & liveness & whenever an instance is required it is returned eventually\\
						 & \namedlabel{res:s:2}{S2} & safety & at every time, there is only one component of type singleton active\\
		\emph{MVC}  	 & \namedlabel{res:m:1}{M1} & liveness & views are eventually updated with user events observed by controllers\\ 
						 & \namedlabel{res:m:2}{M2} & liveness & the data of model and view are eventually in a consistent state\\ 		
		\emph{\broker{}}	 & \namedlabel{res:b:1}{B1} & liveness & every service requires from a client is eventually executed by some server\\ 
						 & \namedlabel{res:b:2}{B2} & liveness & a server requests to registered for a broker eventually results in a registration\\
		\bottomrule
	\end{tabular} 
\end{table}

\subsection{Singleton}
As shown in \tab[0]{tab:res}, we verified two common properties for the \singleton{} pattern. 
Finding \ref{res:s:1} ensures that architectures applying the pattern are guaranteed to eventually activate a singleton if required.
One observation we found during analysis is that the time required to access a singleton component may vary depending on whether it was the first access to the component or a subsequent access. This is because the first access usually requires to newly instantiate the singleton which requires some time.

The second finding \ref{res:s:2} states another characteristic property of the \singleton{} pattern: Each architecture applying the pattern is guaranteed to have at most one active component of type singleton.
During the analysis of this property, we found that there might be different interpretations of the pattern. First, the condition that there can only be one instance, can be interpreted as one instance which is activated every time or as maximal one instance (i. e. none or one).
Another question that arises is whether the changing instance needs to be always the same or whether it is allowed to change over time.

\subsection{MVC}
\tab[1]{tab:res} shows the properties verified for the MVC pattern.
The first property \ref{res:m:1} ensures that whenever an event is registered by the controller, all the views request a new copy of their data from the model.
It is important that the views should be initialized in the idle state instead of the busy state since they need to request the model data at the beginning.

Finding~\ref{res:m:2} ensures consistency of the data from the views with the data from the model. It states that whenever data changes in the model, the corresponding data in the view is eventually updated.

\subsection{Broker}
As shown in \tab[0]{tab:res}, we investigated two properties for the \broker{} pattern.
Finding \ref{res:b:1} states that each service requested by a client is eventually executed by some of the servers. Thereby, the client communicates only with the broker and does not have to know where the service is actually executed.

Finally, finding \ref{res:b:2} ensures that a server which wants to be registered to the broker, is guaranteed to be eventually registered. This is important to guarantee that the services of a registered server can be accessed by the clients.

In order to not complicate the pattern, we did not incorporate a bridge component and stick to one unique broker component.
Our investigation suggests that there are different possibilities to implement the pattern. With our implementation every request is handled for sure, but only because the requesting itself is restricted. In practice, the client, the broker, and the server usually execute in different locations, which is why the information also has to be marshalled and un-marshalled.
\section{Discussion}\label{sec:discussion}
In the following section we summarize our experience with using \mc{} techniques to investigate patterns for dynamic architectures.

In general, we found that pattern specifications are usually underspecified in literature and that many different interpretations are possible where only some of them lead to the desired properties. Consider, for example, the \singleton{} pattern where there is ambiguity in whether the instance has to be active all time or can get deactivated and another instance can get active. We see our approach as helpful for architects to find and tackle these questions.


\paragraph{Scalability}
When it comes to formal methods, scalability is sometimes considered a critical issue. We tried to overcome this problem by raising the level of abstraction of our analyses. By investigating patterns, rather than concrete architectures, we concentrate on the architecturally important aspects and thus reduce complexity of the analyses.

\tab[1]{tab:ltlspec} shows the computational effort to perform our analyses on a standard notebook. The table shows that the computational effort for single as well as for medium complicated patterns (\singleton{} and MVC) is very low ($25$ ms and $30$ ms, respectively).
Note that the \broker{} pattern is actually one of the more complex patterns documented in literature. However, we can see that also for this pattern the computational effort to analysis is rather low ($1000$ ms).

\begin{table}
	\caption{Computational effort.\label{tab:ltlspec}}\small
	\begin{tabular}{rllllll}
		\toprule
		\textbf{Pattern}    & \textbf{Lines of Code} & \textbf{LTL Specs} & \textbf{LTL average} & \textbf{LTL total} & \textbf{\#Variables} & \textbf{\#Components} \\
		\midrule
		\emph{\singleton{}} & $68$                   & $2$                & $25$ ms              & $50$ ms            & $10$                 & $1$                   \\
		\emph{MVC}          & $89$                   & $30$               & $30$ ms              & $1500$ ms          & $15$                 & $4$                   \\
		\emph{\broker{}}    & $309$                  & $6$                & $1000$ ms            & $6000$ ms          & $42$                 & $6$                   \\
		\bottomrule
	\end{tabular}
\end{table}

\noindent
\begin{minipage}{\textwidth}
	\begin{wrapfigure}[25]{r}{0.50\textwidth}
		\vspace{-10pt}
		\centering 
		\includegraphics[width=7cm]{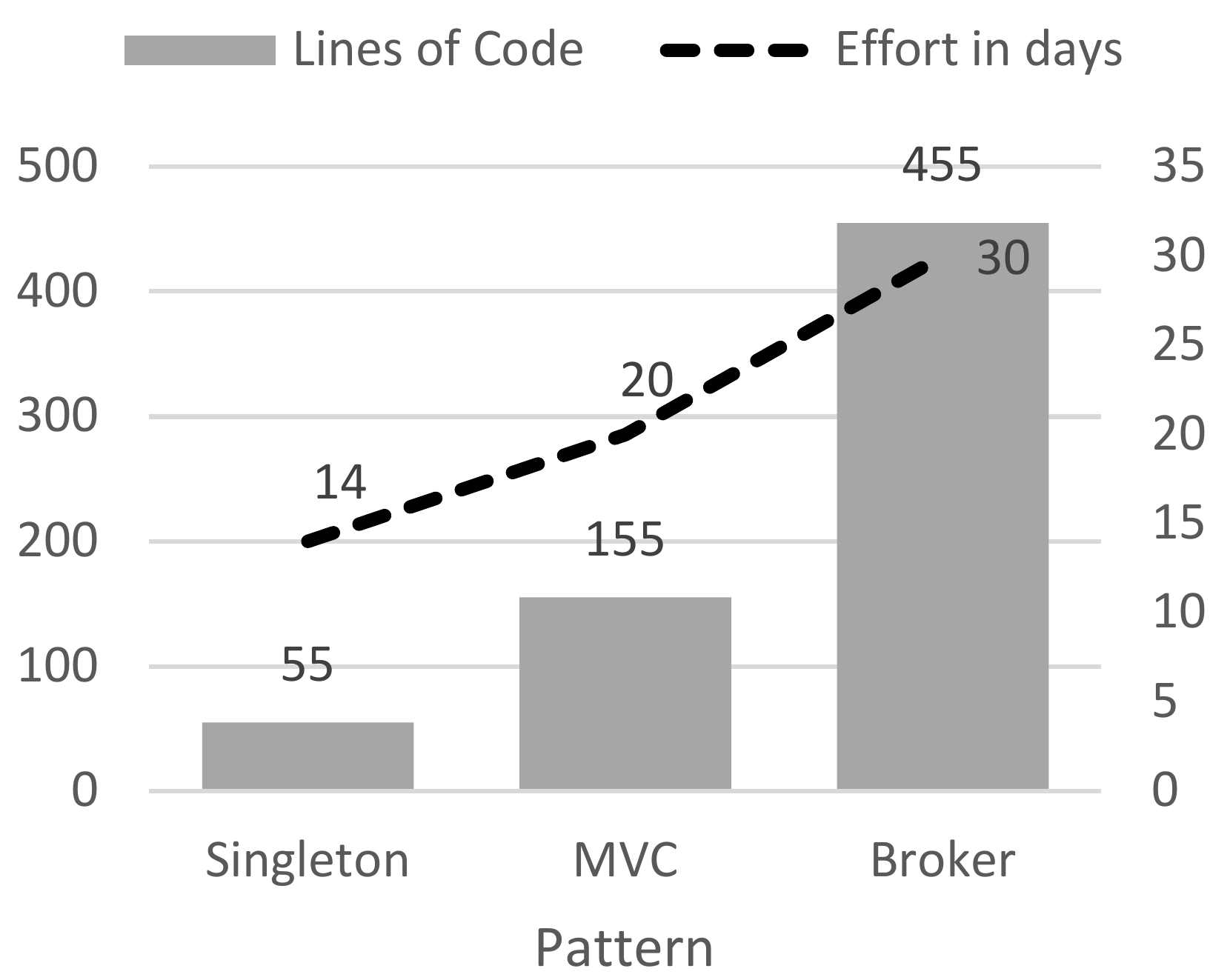}
		\caption{Effort applying the approach.}
		\label{fig:effort} 
	\end{wrapfigure}

	\paragraph{Effort}
	Another problem usually mentioned in formal methods is the required effort.
	\fig[1]{fig:effort} provides an overview of the effort needed to perform the analyses described in the paper\footnote{All analyses were conducted by the same person. The person had a computer science background with no experience of using Model checking techniques so far.}.
	We can see that the \emph{relative} effort strongly decreased with growing experience. Indeed, after some experience, also relatively complex patterns (e.g. the \broker{} pattern) can be analyzed with a reasonable amount of effort (1 PM).
		
	Again we would like to point out the impact of the effort. Since the results are at pattern level, the impact is high since it influences every concrete architecture implemented in a certain pattern.
	\vspace{20pt}
\end{minipage}

\paragraph{Quality attributes.}
A last point which needs to be discussed in more detail regards an important aspect of software architectures in general.
Our approach does actually not provide means to directly specify quality attributes such as performance, availability, etc.
However, as our example shows, it allows us to specify the technical realization of such aspects.
For example, finding~\ref{res:s:2} of the Singleton pattern can be used to ensure an upper bound of active components at each point in time.
This could be actually seen as a possible realization of one aspect of efficiency.
\section{Related Work}\label{sec:rw}
Recently, some approaches emerged which focus on the verification of software architectures and architecture patterns.
They can roughly be classified into two main groups: automatic and interactive verification.

\subsection{Interactive Pattern Verification}
Work in this area applies interactive theorem proving (ITP) to pattern verification.
One example comes from Marmsoler and Gleirscher~\cite{Marmsoler2016} who apply the Isabelle/HOL ITP to investigate patterns for dynamic architectures. While interactive approaches are very expressive, they usually require manual interaction for verification.
Thus, with our approach we actually complement work in this area by providing an alternative to verify certain properties automatically.

\subsection{Automatic Pattern Verification}
Work in this area applies automatic techniques to pattern verification.
One example of work in this area comes from Kim and Garlan~\cite{Kim2006} who apply the Alloy~\cite{Jackson2002} analyzer to automatically verify architectural styles specified in ACME~\cite{Garlan2003}.
A similar approach comes from Wong et al.~\cite{Wong2008} which also applies Alloy to the verification of architecture models.
In contrary to this work, however, both approaches are using the Alloy Analyzer, which is based on SAT-solving whereas we are using \mc{} with focus on LTL formulae.

Another related approach in this area comes from Wirsing et al.~\cite{Wirsing2012} where the authors apply rewriting logic to specify and verify cloud-based architectures. Again, the authors apply rewriting logic whereas the focus of this work was to investigate the feasibility of \mc{} technologies.

Finally, Zhang et al.~\cite{Zhang2012} applied \mc{} techniques to verify architectural styles formulated in Wright\#, an extension of Wright~\cite{Allen1997}.
However, whereas their work focuses on strictly static patterns, in this work we aim to support dynamic architectures, as well.

Indeed, to the best of our knowledge this is the first attempt to apply \mc{} to the verification of patterns for dynamic architectures.
\vspace{-5pt}\section{Conclusion}\label{sec:conc}\vspace{-5pt}
With this paper we report on our experience of applying \mc{} for the \emph{verification} of patterns for dynamic architectures.
To this end, we first describe a $5$-step approach to systematically formalize a pattern and its corresponding guarantees and employ \mc{} to verify the guarantees against the model of the pattern.
Then, we apply the approach to investigate $3$ commonly used patterns: the \singleton{}, the \mvc{}, and the \broker{} pattern. For each pattern, we formalized and verified two characteristic properties.

We found that patterns (also well-known once) are usually underspecified in literature and that many different interpretations are possible where only some of them lead to the promised guarantees.
Thus, we conclude that the proposed approach is useful since it helps to discover such ambiguities.
Moreover, our work shows that the approach is feasible also for relatively complex patterns (such as the \broker{} pattern).
The additional effort is justified by the general nature of the results: each result at pattern level applies to every architecture applying the pattern.

We envisage three possible implications of our results:
\begin{inparaenum}[(i)]
	\item \emph{Analysis of existing patterns}: our approach can be used to specify and analyze existing patterns and uncover ambiguities and flaws in their specification.
	\item \emph{Design of new patterns}: the approach can also be applied to help in the design of new patterns. A pattern can be formally specified and then be verified by the designer without even requiring an implementation thereof.
	\item \emph{Pattern conformance analysis}: the formal pattern specifications may be used to help verifying that an architecture indeed implements a certain pattern. For example, the abstract model of the \singleton{} pattern provided in Ex.~\ref{exmp:model} may be used to check whether a concrete architecture indeed implements this pattern.
	In turn it is guaranteed that the architecture fulfills the desired guarantee provided in Ex.~\ref{exmp:mvc:const}.
\end{inparaenum}
To support these implications, future work is needed in two major areas:
First, the approach should be applied to formalize and analyze further patterns (existing ones as well as new ones).
Then, the results should be incorporated into tools to support in the (possible automatic) verification of pattern conformance.\vspace{-15pt}

\paragraph{Acknowledgments.}
We would like to thank Manfred Broy, Vasileios Koutsoumpas and all the anonymous reviewers for their comments and helpful suggestions on earlier versions of this paper.\vspace{-10pt}
\vspace{-5pt}
\bibliographystyle{eptcs}
\let\oldbibliography\thebibliography
\renewcommand{\thebibliography}[1]{%
	\oldbibliography{#1}%
	\setlength{\itemsep}{0pt}%
}
{\footnotesize
\bibliography{references,additionalRefs}}
\end{document}